# Critical Thermodynamics of Two-Dimensional Systems in the Five-Loop Renormalization-Group Approximation

E. V. Orlov and A. I. Sokolov

*St. Petersburg State University of Electrical Engineering, St. Petersburg, 197376 Russia*
*e-mail: ais2002@mail.ru*

**Abstract**—The paper is devoted to the calculation of renormalization-group (RG) functions in the $O(n)$-symmetry two-dimensional model of the $\lambda\varphi^4$ type in the five-loop approximation and to an analysis of the critical behavior of systems described by this model. Five-loop expansions for the $\beta$ function and the critical indices are determined in bulk theory. They are summed up using the Padé–Borel and Padé–Borel–Le Roy methods, making it possible to optimize the summation procedure and to estimate the accuracy of the obtained numerical values. It is shown that in the Ising ($n = 1$) case, as well as in other cases, the inclusion of the five-loop contribution to the $\beta$ function displaces the coordinate of the Wilson fixed point only insignificantly, leaving it outside the interval formed by the results of computations on lattices; even "spreads" of the error in the renormalization group and lattice estimates do not overlap. This discrepancy is attributed to the effect of the nonanalytic component of the $\beta$ function, which cannot be determined in perturbation theory. A computation of critical indices proves that, although the inclusion of the five-loop terms in the corresponding RG expansion slightly improves the concordance with the exact results, the nonanalytic contributions are apparently also significant in this case.

The renormalization group (RG) method is one of the most important analytical tools applied at present for a theoretical analysis of critical phenomena. It proved to be exceptionally efficient as applied to three-dimensional systems both for determining the quantitative characteristics of the critical behavior and for analyzing qualitative features of phase transitions. The calculation of multiloop RG expansions for an $O(n)$-symmetry model of the $\lambda\varphi^4$ type and their processing with the help of various methods of summation made it possible to obtain highly accurate values of critical indices, critical-amplitude ratios, and renormalized coupling constants [1–9] used as standards for comparing the predictions of the theory with the results of physical and computer experiments.

On the other hand, the advances of the RG method in the theory of phase transitions, which are quite obvious, lack sufficient theoretical substantiation. Indeed, all observables can be presented in this case in the form of a diverging power series in dimensionless renormalized coupling constants which are not small in the critical region. The construction of various iterative procedures on the basis of diverging RG expansions, the best of which exhibit rapid convergence and lead to matching numerical results, simplifies the problem, but naturally does not solve it. In this situation, alternative methods of verification of the reliability and efficiency of the RG method (primarily, its testing on certain exactly solvable models) become especially significant.

The two-dimensional Ising model, describing critical phenomena in a number of real physical objects, is a well-known example of an exactly solvable model of phase transitions. It is generally accepted that in the critical region this model is thermodynamically equivalent to the two-dimensional scalar theory of the $\lambda\varphi^4$ type, and hence its critical behavior can be analyzed using the RG method in analogy with three-dimensional systems. Quite recently, exact values of asymptotic critical indices were also determined for a family of two-dimensional models corresponding to conformal-invariant theories [10–12]. These models are characterized by $n$-component order parameters with nonintegral values of $n$, forming an infinite sequence converging to the point $n = 2$. They form a natural basis for further testing of the RG method in phase-transition theory.

Another circumstance stimulating the study of two-dimensional models by using the RG technique is that not all of the universal parameters characterizing their critical behavior are known or can be determined from the exact solutions obtained. As a matter of fact, these solutions are valid either in a zero field or only at the critical point, and hence they cannot be used, for example, to find the critical index of corrections to scaling $\omega$ or the renormalized dimensionless coupling constants $g_{2n}$ appearing in the equation of state (a detailed analysis of questions associated with this problem can be found, for example, in [13, 14]). At the same time, the recent renormalization-group calculation of the universal value of the vertex $g_6$ for the two-dimensional Ising model [15] and a comparison of the result obtained with the lattice analogs [16, 17] proved that the RG method can be sufficiently efficient in such cases.

It should be noted that in the physical-dimensionality space the field-theoretical RG method was applied for the first time to the two-dimensional Ising model more than two decades ago. In the classical work [1], the RG functions of a two-dimensional model of the $\lambda\varphi^4$ type with $n = 1$ were calculated in the four-loop approximation. The summation of the obtained RG expansions by using the Padé–Borel–Le Roy method led to estimates of "large" critical indices $\gamma$ and $\nu$, which were found to be in accord with the Onsager values. However, the values 0.16, 0.08, and 0.06 obtained for "small" indices $\alpha$, $\eta$, and $\beta$ differ significantly from the exact values (0, 1/4, and 1/8). In addition, the comparatively small number of terms in the RG series and their stronger convergence than in the 3D case deteriorated the accuracy of the numerical results: the values of the corresponding "spreads" were found to be between ±0.2 and ±0.6 [1]. The application of more complicated summation methods [2] based on the Borel–Le Roy transformations and on the conformal mapping technique slightly improved the situation; the estimates $\alpha = 0.06 \pm 0.24$, $\eta = 0.13 \pm 0.07$, and $\beta = 0.08 \pm 0.26$ were obtained for small indices, but their difference from the exact values still remained too large, and the accuracy was quite low.

This work aims at the calculation of the RG functions in a two-dimensional $O(n)$-symmetry model of the $\lambda\varphi^4$ type in the five-loop approximation. These functions will be determined for an arbitrary $n$. Summation of the RG expansions will allow us to determine the coordinate of a nontrivial fixed point and the critical indices for the cases $n = 1$ and 0, which are interesting from the physical point of view and correspond to layered Ising ferromagnets and polymers, as well as for an exactly solvable model with $n = -1$. The article has the following structure. Section 1 is devoted to determining the RG expansions for the $\beta$ function and critical indices. In Section 2, the Padé–Borel–Le Roy method of summation of the expansion of the $\beta$ function is used to calculate the coordinate of a nontrivial (Wilson) fixed point $g_4^*$ and the index $\omega$ of corrections to the scaling. Padé's approximants of several different types are used in this case, and the summation procedure is optimized. Section 3 contains an analysis of asymptotic critical indices for the above values of $n$, a comparison of the numerical results with the exact values and with the data of calculations on lattices, and a discussion of the efficiency of the field-theoretical RG method as applied to problems of the type under consideration.

## 1. FIVE-LOOP EXPANSIONS FOR THE $\beta$ FUNCTION AND CRITICAL INDICES

Thus, the Hamiltonian of the model under consideration has the form

$$H = \int d^2x \left[ \frac{1}{2}(m_0^2 \varphi_\alpha^2 + (\nabla \varphi_\alpha)^2) + \frac{\lambda}{24}(\varphi_\alpha^2)^2 \right], \quad (1)$$

where $\varphi_a$ is the real $n$-component vector field, the square of the "bare mass" $m_0^2$ is proportional to $T - T_c^{(0)}$, and $T_c^{(0)}$ is the phase transition temperature disregarding order parameter fluctuations.

We will calculate the $\beta$ function and critical indices in the framework of bulk theory. In Green's function, the vertex part and the total three-leg vertex are assumed to be normalized for zero external momenta according to the conventional procedure

$$G_R^{-1}(0, m, g_4) = m^2, \quad \left.\frac{\partial G_R^{-1}(p, m, g_4)}{\partial p^2}\right|_{p^2 = 0} = 1, \quad (2)$$
$$\Gamma_R(0, 0, 0, m, g) = m^2 g_4, \quad \Gamma_R^{(1, 2)}(0, 0, m, g_4) = 1.$$

Since four-loop expansions for the $\beta$ function and critical indices for $n = 1$ are known [1], we must obtain the corresponding series for arbitrary $n$ and then calculate the five-loop contributions. The solution of the first problem is not complicated by any difficulties, since the combinatorial factors, tensor convolutions, and the numerical values of integrals of all one-, two-, three-, and four-loop Feynman diagrams were determined earlier [18]. Conversely, the integrals corresponding to five-loop vertex and mass diagrams have not been calculated for the 2D case and will be calculated here for the first time. Without going into the details of this computation, we consider the most significant aspects of the analysis.

The five-loop contribution to the total four-leg vertex is specified by the sum of 124 topologically different diagrams, which are compiled in [18]. Twenty-seven diagrams have a trivial structure in the sense that their integrals are the products of the integrals of lower-order diagrams. Several dozens of diagrams correspond to integrals which can easily be evaluated with the help of a computer, since they can be reduced to single or double integrals. However, the calculation of triple and more complex integrals cannot be carried out using standard packages and requires the application of appropriate programs, which were specially developed for this purpose. The calculation of 31 five-fold and three seven-fold integrals was the most time-consuming. The latter were evaluated to within four decimal places, but the relative total contribution of these three diagrams to the five-loop term was approximately equal to 2.5%. Because the remaining diagrams were calculated with errors that were several orders of magnitude lower, the accuracy of the final result was better than to five decimal places. As a matter of fact, the accuracy proved to be still higher, since the experience of performing operations with our programs shows that the error in such computations is actually an order of magnitude smaller than that declared by the corresponding option. For this reason, we will henceforth write the five-loop contribution to the $\beta$ function to within six decimal places.

Thus, the expansion of the β function in the model with the Hamiltonian (1) has the form

$$\frac{\beta(g)}{2} = -g + g^2$$

$$-\frac{g^3}{(n+8)^2}(10.33501055n + 47.67505273)$$

$$+\frac{g^4}{(n+8)^3}(5.0002759n^2$$

$$+ 149.1518586n + 524.3766023)$$

$$-\frac{g^5}{(n+8)^4}(0.08884291n^3 + 179.6975910n^2 \quad (3)$$

$$+ 2611.154798n + 7591.108694)$$

$$+\frac{g^6}{(n+8)^5}(-0.00407946n^4 + 80.3096n^3$$

$$+ 5253.56n^2 + 53218.6n + 133972).$$

The calculation of five-loop RG expansions for critical indices also required the computation of quite a family of multiple integrals, which turned out to be more complicated than in the case of the β function. Eventually, the following expressions were obtained for the indices γ and η:

$$\gamma^{-1} = 1 - \frac{n+2}{n+8}g + \frac{g^2}{(n+8)^2}(n+2)3.3756289$$

$$-\frac{g^3}{(n+8)^3}(4.6618848n^2$$

$$+ 34.41848329n + 50.18942749)$$

$$+\frac{g^4}{(n+8)^4}(0.31899304n^3 + 71.70330240n^2 \quad (4)$$

$$+ 429.4244948n + 574.5877236)$$

$$-\frac{g^5}{(n+8)^5}(0.0938051n^4 + 85.4975n^3$$

$$+ 1812.19n^2 + 8453.70n + 10341.1).$$

$$\eta = \frac{g^2}{(n+8)^2}(n+2)0.91708597$$

$$-\frac{g^3}{(n+8)^3}(n+2)0.05460898$$

$$+\frac{g^4}{(n+8)^4}(-0.09268446n^3 + 4.05641051n^2 \quad (5)$$

$$+ 29.2511668n + 41.5352155)$$

$$-\frac{g^5}{(n+8)^5}(0.0709196n^4 + 1.05240n^3$$

$$+ 57.7615n^2 + 325.329n + 426.896).$$

Here, as in the previous publications [1–3], instead of the renormalized coupling constant $g_4$, we used the dimensionless invariant charge proportional to it:

$$g = \frac{n+8}{24\pi}g_4, \quad (6)$$

which, in contrast to $g_4$, does not tend to zero as $n \longrightarrow \infty$, but attains a finite value equal to unity.

## 2. THE COORDINATE OF WILSON'S FIXED POINT IN THE FIVE-LOOP APPROXIMATION

The values of indices and other universal parameters characterizing a phase transition are determined by the coordinate of Wilson's fixed point $g^*$, which is a non-trivial solution to the equation $\beta(g) = 0$. Like other series of the renormalized perturbation theory, the expansion obtained by us for $\beta(g)$ is asymptotic; in order to find $g^*$, the series in Eq. (3) must be reduced to a convergent one, i.e., subjected to a rearrangement of its terms. This is usually done using the Borel–Le Roy transformation

$$f(x) = \sum_{i=0}^{\infty} c_i x^i = \int_0^{\infty} e^{-t}t^b F(xt)dt,$$

$$F(y) = \sum_{i=0}^{\infty} \frac{c_i}{(i+b)!}y^i. \quad (7)$$

In order to evaluate the integral in Eq. (7), the Borel transform $F(y)$ of the required function must be continued analytically beyond the convergence range. To this end, we can use Padé's approximants [L/M], which are the ratios of polynomials $P_L(y)$ and $Q_M(y)$ of the Lth and Mth degree, respectively, whose coefficients are defined unambiguously if the sum $L + M + 1$ coincides with the number of the known terms of the series, and $Q_M(0) = 1$. It was found that the best approximating properties are observed for the diagonal Padé approximants, for which $L = M$, or approximants close too them (see, for example, [19]). However, the number of roots, i.e., the number of approximant poles in the complex plane, increases with the degree of the denominator $M$. If at least some of these poles are close to the real semi axis $y > 0$, or, which is still worse, lie on this semi axis, the corresponding approximant becomes unsuitable for the summation of the series. In actual practice, this considerably limits the degree of the denominator from above and narrows the choice of admissible approximants. On the other hand, the presence of the adjustable parameter $b$ in the Borel–Le Roy transfor-mation makes it possible to optimize the summation procedure by attaining the most rapid convergence of the iterative process.

**Table 1.** Coordinate of Wilson's fixed point for models with $n = 1, 0,$ and $-1$ as calculated in four successive RG approximations and the resultant five-loop estimates of $g^*(n)$

| $n$ | [1/1] | [2/1] | [2/2] | [3/2] | $g^*$, 5-loop |
|---|---|---|---|---|---|
| 1 | 2.4246 | 1.7508 | 1.8453 | 1.8286 | $1.837 \pm 0.03$ |
| 0 | 2.5431 | 1.7587 | 1.8743 | 1.8402 | $1.86 \pm 0.04$ |
| $-1$ | 2.6178 | 1.7353 | 1.8758 | 1.8278 | $1.85 \pm 0.05$ |

Taking into account what has been said above, we initially chose the following procedure for calculating the coordinate of the nontrivial fixed point $g^*$ [6]. For each $n$, a nontrivial root of the equation $\beta(g) = 0$ was determined in four successive (two-, three-, four-, and five-loop) approximations, and the analytic continuation of the Borel transforms of the $\beta$ function was carried out using symmetric or nearly symmetric Padé approximants: [1/1], [2/1], [2/2], and [3/2]. The parameter $b$ was varied over wide limits and chosen such that the numerical results specified by higher (four- and five-loop) approximations coincided; i.e., the most rapid convergence of the iterative procedure was ensured. Unfortunately, the optimal values of $b$ corresponding to the most rapid convergence proved to be quite close to the threshold values, i.e., the values beyond which working approximants acquire poles for positive values of $y$. In actual practice, this circumstance significantly affects the accuracy of the results obtained.

In order to avoid the problem of poles, we did not use the variation of $b$ and confined the analysis (for small $n$) to a fixed value of this parameter (equal to zero), which corresponds to the Padé–Borel summation method. In this case, all the Padé approximants listed above are free from "hazardous" poles, and the iterative procedure converges quite rapidly. The results of computations for $n = 1, 0,$ and $-1$ (exactly solvable models) given in Table 1 clearly illustrate the situation. It can be seen that the application of the approximants [1/1], [2/1], [2/2], and [3/2] for analytic continuation of the Borel transform gives estimates for $g^*$ that rapidly approach asymptotic values; the process of attaining the asymptotic form has the form of damped oscillations. The presence of oscillations appears quite natural, since the series for the $\beta$ function is alternating, and their damping reflects the Borel summability of the RG expansion. Consequently, it can be concluded that the asymptotic value of $g^*$ must lie between the four- and five-loop estimates, and it is natural to take their half-sum as the final result. For example, having obtained $g^* = 1.8453$ and $1.8286$ in the four- and five-loop approximations of the two-dimensional Ising model, respectively, we assume that $g^* = 1.837$ is the most probable coordinate of Wilson's fixed point. The estimates of $g^*$ for other values of $n$ are given in the last column of Table 1 and in the upper row of Table 2.

The accuracy of determining the coordinate of Wilson's fixed point was estimated as follows. We varied the parameter $b$ from 0 to 10, i.e., over wide limits, and traced the ensuing variation of $g^*$ obtained by averaging the four- and five-loop results. The range of variation of this average value was taken as the error in determining the numerical value of $g^*$. The estimate of error obtained in this way is quite conservative since it exceeds considerably (at least by a factor of two) the difference between the averaged and the five-loop val-

**Table 2.** Coordinate of Wilson's fixed point $g^*$ and the critical index $\omega$ for $-1 \geq n \geq 32$ in the five-loop renormalization-group approximation

| $n$ | $-1$ | 0 | 1 | 2 | 3 | 4 | 8 | 16 | 32 |
|---|---|---|---|---|---|---|---|---|---|
| | | | | $g^*$ | | | | | |
| RG, 5-loop | 1.85(5) | 1.86(4) | 1.837(30) | 1.80(3) | 1.75(2) | 1.70(2) | 1.52(1) | 1.313(3) | 1.170(2) |
| | | | | | | ($b=1$) | ($b=1$) | ([4/1], [3/1]) | ([4/1], [3/1]) |
| HT exp [22, 23] | | 1.679(3) | 1.754(1) | 1.81(1) | 1.724(9) | 1.655(16) | | | |
| MC [25, 29] | | | 1.71(12) | 1.76(3) | 1.73(3) | | | | |
| SC [24] | 1.473(8) | 1.673(8) | 1.746(8) | 1.81(2) | 1.73(4) | | | | |
| $\epsilon$-exp [23] | | 1.69(7) | 1.75(5) | 1.79(3) | 1.72(2) | 1.64(2) | 1.45(2) | 1.28(1) | 1.16(1) |
| $1/n$-exp [23] | | | | | 1.758 | 1.698 | 1.479 | 1.283 | 1.154 |
| | | | | $\omega$ | | | | | |
| RG, 5-loop | 1.32(4) | 1.31(3) | 1.31(3) | 1.32(3) | 1.33(2) | 1.37(3) | 1.50(2) | 1.70(1) | 1.85(2) |

Note: The values of $g^*$ extracted from high-temperature (HT) expansions and strong-coupling (SC) expansions, as well as those calculated by the Monte Carlo (MC) method, obtained by the processing of $\epsilon$ expansion for $g^*$ ($\epsilon$-exp), and specified by the corresponding $1/n$ expansion ($1/n$-exp), are given for comparison.

ues of $g^*$. This allows us to treat this estimate as quite realistic.

Before using the values obtained for determining critical indices, it would be interesting to compare them with the values of $g^*$ reported earlier in other publications. The coordinate of the fixed point for the two-dimensional Ising model was determined by the RG method in the physical-dimensionality space [1, 2, 20] from an analysis of high-temperature expansions [16, 21–23] with the help of the $\epsilon$-expansion technique [23], the Monte Carlo method [25], and the strong-coupling method [24, 26]. The numerical value of $g^*$ was obtained from the results of calculation on lattices with the help of the relation

$$\chi_4 = \left.\frac{\partial^3 M}{\partial H^3}\right|_{H=0} = -\chi^2 m^{-2} g_4, \quad (8)$$

connecting the renormalized coupling constant $g_4 = (8\pi/3)g$ with the nonlinear $\chi_4$ and conventional $\chi$ susceptibilities of the system in the critical region. The summation of four-loop RG expansions by the Padé–Borel ([3/1] approximant) and Padé–Borel–Le Roy methods, as well as by using the conformal mapping technique, led to the following estimates: $g^* = 1.88$ [20], $1.8 \pm 0.3$ [1], and $1.85 \pm 0.1$ [2], respectively. The processing of high-temperature expansions, as well as of strong-coupling expansions, made it possible to obtain close values characterized by a high expected accuracy: $g^* = 1.751$ [21], $1.7547 \pm 0.002$ [16], $1.7538 \pm 0.0005$ [22], and $1.746 \pm 0.008$ [24]. The summation of the $\epsilon$-expansion for the renormalized coupling constant for $\epsilon = 2$ by using information on exact values of $g^*$ for low-dimensional models ($D = 1$ and $0$) led to the estimates of $g^* = 1.79 \pm 0.05$ and $1.75 \pm 0.05$ [23]. Finally, the Monte Carlo method and direct summation of the strong-coupling expansion for the $\beta$-function resulted in $g^* = 1.71 \pm 0.12$ [25] and $1.76$ [26].

A comparison of these numbers with one another and with our result $g^* = 1.837 \pm 0.03$ leads to the following important conclusions. First, the lattice estimates of $g^*$, grouped around the value of $g^* = 1.75$, noticeably differ from their analogs obtained by the RG method in the four-loop approximation. Second, the inclusion of the five-loop contribution to the $\beta$-function, which improves considerably the expected accuracy in the determination of $g^*$, leads only to an insignificant displacement of the coordinate of the fixed point, leaving it outside the interval containing the results of the calculations on lattices. Moreover, even the spreads in the errors of the RG and lattice estimates do not overlap in the five-loop approximation. The reasons behind this discrepancy can be associated with the insufficient length of the available RG expansions and the slower convergence of iterations than in the $D = 3$ case on the one hand, and, on the other hand, with the presence of nonanalytic contributions to the RG functions, which cannot be determined from perturbation theory.

The first reason does not appear to be likely. Indeed, the series for the $\beta$ function is alternating, and hence the dependence of $g^*$ on the approximation order is oscillating by nature. This means that the inclusion of the six-loop term in the expansion of $\beta(g)$ leads to an increase in $g^*$, i.e., to a larger difference between the RG result and its lattice analogs. The perturbative contributions of higher orders might slightly reduce the six-loop estimate, but the value of $g^*$ at any rate remains larger than that obtained in the five-loop approximation in view of the convergence of the iterative procedure. Consequently, the divergence under consideration cannot be eliminated in perturbation theory.

It is natural to attribute this divergence to the effect of the nonanalytic component of the $\beta$ function. It is well known that field-theoretical functions must have singularities [27] (Dyson theorem) at the point $g = 0$, near which the weak-coupling expansions are constructed. In a theory of the $\lambda\varphi^4$ type, Wilson's fixed point itself can be singular for the $\beta$ function [23, 28]. Numerous calculations made in recent decades show that the nonanalyticity of RG functions virtually does not affect the accuracy of determining the critical indices and other universal quantities characterizing the critical behavior of 3D systems. However, the role of singular terms must increase with decreasing dimensionality. The results obtained can be regarded as a convincing demonstration of the fact that the influence of nonanalytic terms for 2D objects is no longer negligibly small.

This conclusion is valid not only for the Ising model. For $n \neq 1$, the field-theoretical RG method also leads to estimates of $g^*$, which differ significantly from the numbers obtained by lattice calculations. For example, for $n = 0$, the method of high-temperature expansions and the RG analysis in the five-loop approximation give $g^* = 1.679 \pm 0.003$ [24] and $1.86 \pm 0.04$, respectively (see Table 1). With increasing $n$, the difference between the lattice and the RG estimates of the coupling constant decreases but remains comparable with the errors in determining $g^*$ or exceeds them. This is clearly illustrated in Table 2, containing, in addition to five-loop RG estimates, the values of $g^*$ for various values of $n$ obtained from high-temperature expansions [22, 23] (row 2), strong-coupling expansions [24] (row 4), and the Monte Carlo method [25, 29] (row 3), as well as those leading to the $\epsilon$ expansion summed up taking into account the available exact values of $g^*$ for $D = 1$ and $0$ (row 5) and the $1/n$ expansion (row 6); the latter are borrowed from [23]. (In order to avoid a misunderstanding, we note that the models with $n > 1$ are considered here exclusively for testing the RG method rather than for describing the thermodynamics of real degenerate 2D systems, in which ferromagnetic transitions are known to be absent.) In order to determine $g^*$ for $n = 16$ and $32$, Padé approximants [4/1] and [3/1] were used, since the values of the coupling constant obtained on their basis depend on the parameter $b$ only

**Table 3.** Critical indices for models with $n = 1, 0$, and $-1$ determined by the Padé–Borel summation of five-loop RG expansions

| $n$ | Method | $g^*$ | $\gamma$ | $\eta$ | $\nu$ | $\alpha$ | $\beta$ |
|---|---|---|---|---|---|---|---|
| 1 | RG | 1.837 | 1.790 | 0.146 | 0.966 | 0.068 | 0.071 |
|   |    | 1.754 (HT) | 1.739 | 0.131 | 0.931 | 0.139 | 0.061 |
|   | Exact |  | 7/4 | 1/4 | 1 | 0 | 1/8 |
|   |    |  | (1.75) | (0.25) |  |  | (0.125) |
| 0 | RG | 1.86 | 1.449 | 0.128 | 0.774 | 0.452 | 0.049 |
|   |    | 1.679 (HT) | 1.402 | 0.101 | 0.738 | 0.524 | 0.037 |
|   | Exact |  | 43/32 | 5/24 | 3/4 | 1/2 | 5/64 |
|   |    |  | (1.34375) | (0.20833) | (0.75) | (0.5) | (0.078125) |
| $-1$ | RG | 1.85 | 1.184 | 0.082 | 0.617 | 0.765 | 0.025 |
|   |    | 1.473 (SC) | 1.155 | 0.049 | 0.592 | 0.816 | 0.014 |
|   | Exact |  | 37/32 | 3/20 | 5/8 | 3/4 | 3/64 |
|   |    |  | (1.15625) | (0.15) | (0.625) | (0.75) | (0.046875) |

Note: The numerical values of these indices are also given for comparison.

slightly, and the approximants [3/2] and [2/2] become inapplicable for large values of $n$ in view of the emergence of "hazardous" poles. It can be seen from Table 2 that the RG and lattice estimates of $g^*$ are close to each other only for $n = 2$ and 3. This closeness, however, is accidental and does not change the conclusion concerning the systematic divergence of the field-theoretical and lattice estimates of the coordinate of the fixed point.

Apart from the numerical values of the coordinate of Wilson's fixed point, Table 2 also contains our estimates of the critical index $\omega = d\beta(g^*)/dg$ determining the temperature dependences of scaling corrections. The index $\omega$ was determined by numerical differentiation of the function $\beta(g)$ specified by the RG expansion summed up according to the Padé–Borel method (approximants [3/2] and [2/2]), and the error was taken as half the difference between the five- and four-loop estimates of this index. Since the diagonal and close-to-diagonal approximants acquire "hazardous" poles for $b = 0$ with increasing $n$ (see above), the shift parameter for determining the index $\omega$ was taken as 1 for $n = 4$ and 8, while for $n = 16$ and 32, the approximants [4/1] and [3/1] were used (for $b = 0$).

## 3. CRITICAL INDICES: DISCUSSION OF RESULTS

Let us now determine the numerical values of the critical indices. It is well known that RG expansions for different indices differ considerably in their structure. For example, series (4) for $\gamma^{-1}$ is alternating and is characterized by a regular behavior of the coefficients, which does not apply to the RG expansions of $\gamma$ and $\nu$. In order to ensure the highest rate of convergence of the iterative procedure, we carried out the Padé–Borel–Le Roy summation of the series for $\gamma^{-1}$ and $\eta$, while the remaining critical indices were determined with the help of the well-known scaling relations. In order to verify the self-consistency of the obtained numerical results and to estimate their accuracy, we also calculated the indices

$$\eta^{(2)} = \frac{1}{\nu} + \eta - 2, \quad \eta^{(4)} = \frac{1}{\nu} - 2, \qquad (9)$$

whose RG expansions have a regular structure. Since the RG method gives the coordinate of Wilson's fixed point which differs noticeably from the results of lattice calculations (see above), the critical indices were determined using both the renormalization-group and the lattice values of $g^*$. This enabled us to determine the values of $g^*$ ensuring the closeness of RG estimates of critical indices to the exact values and to ascertain the extent of the sensitivity of these results to the value of $g^*$. While processing the RG expansion for $\gamma^{-1}$, we used the approximant [3/2], while the series for $\eta^{(i)}$ and $\eta$, which start from the first- and second-order terms in $g$, respectively, were summed up (after factoring out common multipliers) with the help of approximants [2/2] and [2/1]. The values of critical indices determined in this way were found to be weakly dependent on the parameter $b$, which can obviously be explained by the high symmetry of the approximants used.

The numerical results obtained for the models with $n = 1, 0$, and $-1$ for $b = 0$ are presented in Table 3. Although the values of critical indices are given in this table to the third decimal place, the actual accuracy of RG estimates is much lower. An estimate of its value can be obtained by calculating the index $\eta$ in two different ways: summing up directly the series in Eq. (5) for this index or determining $\eta$ as the difference of the series summed up for $\eta^{(2)}$ and $\eta^{(4)}$. In the Ising case, the value of $\eta$ determined by the second method is equal to 0.093, i.e., differs from the direct estimate by 0.053; for

$n = 1$, this difference is 0.028. Although the attained accuracy is quite low, it nevertheless makes it possible to characterize the situation quite definitely. The inclusion of five-loop terms in RG expansions obviously leads to a certain decrease in the difference between the renormalization-group estimates and the exact values of critical indices. At the same time, this does not solve the problem of small indices, for which the discrepancy between the predictions of the RG method and the exact values remain on the order of the indices themselves. This conclusion does not depend on whether the values of $g^*$ used for determining critical indices were obtained by the RG method or from the high-temperature expansions ($n = 1, 0$) and the strong-coupling expansion ($n = -1$).

Will the inclusion of the next terms in the RG expansion of critical indices change the situation? In all probability, it will not. Indeed, the series for $\gamma^{-1}$ and $\eta$, as well as for the indices $\eta^{(2)}$ and $\eta^{(4)}$, are alternating, which leads to oscillating dependences of the numerical values of these indices on the approximation order. Since the five-loop estimates of critical indices are closer to the exact values than the four-loop estimates, the addition to the six-loop contributions must deteriorate (at least to a small degree) the quality of the RG estimates. This means that the discrepancy under consideration cannot be eliminated in perturbation theory. It can only be assumed that it originates from nonanalytic contributions to the indices. It was proved that these contributions for 2D models are significant.

In conclusion, let us consider the results of calculations of the critical index $\omega$. It is well known that the true value of the index of scaling corrections in the two-dimensional Ising model remains disputable. The first RG computations in two dimensions (four-loop approximation) gave values of $\omega$ close to 1.3 [1, 2]. The summation of the $\epsilon$ expansion for $\epsilon = 2$ led to the estimate $\omega = 1.6 \pm 0.2$ [30]. This value is in good agreement with the predictions of the conformal-invariant theory, according to which $\omega = 4/3$ in the two-dimensional Ising model [31], and with the results of analysis of high-temperature expansions, according to which $\omega = 1.35 \pm 0.25$ [32]. On the other hand, all the above values contradict the results of exact calculations of the principal singular and correction terms for the susceptibility of the two-dimensional Ising model, which give $\omega = 1$ [33]. Moreover, it was found recently that the two-dimensional Ising model belonging to the family of conformal-invariant theories for which $\omega = 4/m$, with $m = 1, 2, 3, \ldots$, occupies a special place in this family: the index $\omega$ for this model must be equal to 2 [34] and not to 4/3 ($m = 3$ in the Ising model). The only perturbative estimate close to $\omega = 2$ was obtained from an analysis of the strong-coupling expansion for the $\beta$ functions, according to which $\omega = 1.88$ [26]. On the contrary, it can be seen from Table 2 that the results of our calculations confirm the conclusion about the closeness of the index $\omega$ to 4/3 in the two-dimensional Ising model. The inclusion of the five-loop term in $\beta(g)$ made it possible to improve the accuracy of the estimates for $\omega$ as compared to the four-loop approximation and, accordingly, to make this conclusion more definite.

## ACKNOWLEDGMENTS

We are grateful to B.N. Shalaev for numerous discussions of the critical thermodynamics of two-dimensional systems and for helpful suggestions.

This research was supported by the Ministry of Higher Education of the Russian Federation, grant no. 97-14.2-16, the International Science Foundation, A.I.S., grant no. p99-943, and the administration of St. Petersburg, E.V.O., grant no. 298496.